\documentclass[manuscript, review=false, screen=true]{acmart}

\usepackage{todonotes}
\usepackage{placeins}
\usepackage{setspace}
\usepackage{tcolorbox}

\newcommand{\affiliateungp}[0]{
    \affiliation{
      \institution{UN Global Pulse}
    }
    }
    
\newcommand{\affiliateunhcr}[0]{
    \affiliation{
      \institution{UNHCR Innovation}
    }
    }
    
\newcommand{\affiliateunhcrbr}[0]{
    \affiliation{
      \institution{UNHCR Brazil}
    }
    }
    
\newcommand{\affiliateunhcrgds}[0]{
    \affiliation{
      \institution{UNHCR Global Data Service}
    }
    }
    
\newcommand{\affiliateunhcrasia}[0]{
    \affiliation{%
      \institution{UNHCR Asia Bureau}
    }
    }
    
\newcommand{\affiliategatech}[0]{
    \affiliation{
        \institution{Georgia Institute of Technology}
    }
    }

\setcopyright{acmcopyright}
\copyrightyear{2022}
\acmYear{2022}
\acmDOI{}
\acmConference[3rd KDD Workshop on Data-driven Humanitarian Mapping]{August 15, 2022}{Washington, DC}{USA}
\acmBooktitle{Proceedings of the 3rd KDD Workshop on Data-driven Humanitarian Mapping, August 15, 2022, Washington, DC USA} 

\begin{document}

\title[Population Movements under Uncertainty]{Modeling Population Movements under Uncertainty at the Border in Humanitarian Crises: {A Situational Analysis Tool}}

\author{Arturo De Nieves Guti\'errez de Rubalcava} \affiliateunhcrbr
\author{Oscar S\'anchez Pi\~neiro} \affiliateunhcrbr
\author{Rebeca Moreno Jim\'enez} \affiliateunhcr
\author{Joseph Aylett-Bullock} \affiliateungp
\author{Azra Ismail} \affiliategatech \affiliateungp
\author{Sofia Kyriazi} \affiliateunhcr
\author{Catherine Schneider} \affiliateunhcr
\author{Fred Sekidde} \affiliateunhcrbr
\author{Giulia Del Panta} \affiliateunhcrgds
\author{Chao Huang} \affiliateunhcrasia
\author{Vanessa Maign\'e} \affiliateungp
\author{Miguel Luengo-Oroz} \affiliateungp
\author{Katherine Hoffmann Pham} \affiliateungp

\renewcommand{\shortauthors}{De Nieves, et al.}

\begin{abstract}
Humanitarian agencies must be prepared to mobilize quickly in response to complex emergencies, and their effectiveness depends on their ability to identify, anticipate, and prepare for future needs. These are typically highly uncertain situations in which predictive modeling tools can be useful but challenging to build. To better understand the need for humanitarian support -- including shelter and assistance -- and strengthen contingency planning and protection efforts for displaced populations, we present a {situational analysis tool to help anticipate} the number of migrants and forcibly displaced persons that {will} cross 
a border in a humanitarian crisis. The {tool} consists of: (i) indicators of potential intent to move drawn from traditional and big data sources; (ii) predictive models for forecasting possible future movements; and (iii) a {simulation of} border crossings and shelter capacity requirements under different conditions. This tool has been specifically adapted to contingency planning in settings of high uncertainty, with an application to the Brazil-Venezuela border during the COVID-19 pandemic.
\end{abstract}

\maketitle
\section{Introduction}
Humanitarian agencies frequently need to respond to complex emergencies with little time to prepare and incomplete information. In this paper, we describe the development of {a situational analysis tool} specifically designed to help the UN Refugee Agency (UNHCR) model, estimate, and anticipate cross-border flows of migrant and forcibly displaced populations in highly complex {and uncertain} contexts, such as in the aftermath of COVID-related border closures. Predictive models of migration and forced displacement have been developed for specific settings \cite{ suleimenova_generalized_2017,earney_pioneering_2019, huynh_forecasting_2020,team_elva_forecasting_2021,hoffmann_pham_predictive_2022, hegre_forecasting_2022} as well as at a global scale \cite{danish_refugee_council_global_2021, danish_refugee_council_global_2022}. More generally, a range of approaches is available, from short-term early warning systems to mid-term model- or survey-based forecasts and long-term foresight exercises \cite{sohst_forecasting_2020}. 

Our {tool} builds on prior work and is distinct from existing efforts in three ways. First, rather than focusing solely on predictive models of movement, we present a suite of interlinked {tools} that includes efforts at description of current conditions (i.e., nowcasting) and simulation of potential future scenarios as well. Second, given the challenges of building predictive models during the COVID-19 pandemic, the {tool} has been designed with a focus on highlighting, illustrating, and helping understand \textit{uncertainty}, which is a defining characteristic of humanitarian emergencies that makes precise prediction especially difficult. Third, the {tool} and its outputs have been co-designed through close collaboration with operational teams on the ground at UNHCR over the course of over a year and a half, and it is currently deployed in order to inform ongoing operations. 

 We developed the {tool} in the context of the crisis at the Venezuela-Brazil border during the COVID-19 pandemic. Venezuelans represent one of largest populations of concern to UNHCR, with over 330,000 Venezuelans estimated to be in Brazil as of 2022 \cite{iom_subcomite_2022}. Many Venezuelans regularly cross the border between Brazil and Venezuela for protection, economic, or family reasons \cite{unhcr_venezuela_2022, r4v_monitoramento_2020}, but on March 18, 2020, the border was closed in response to the COVID-19 pandemic. As the Brazilian government considered reopening the border in 2021, there was a clear need for predictive analytics and simulation tools to estimate the number of {potential} future crossings and the additional shelter capacity that would be necessary to host incoming Venezuelans.
We have continued to develop the tool even as the border reopened, with the goal of using it to inform continuous operations in a number of scenarios and contexts. 

\section{{Overview of the situational analysis tool}}

\begin{figure}
    \centering
    \includegraphics[width=.8\linewidth]{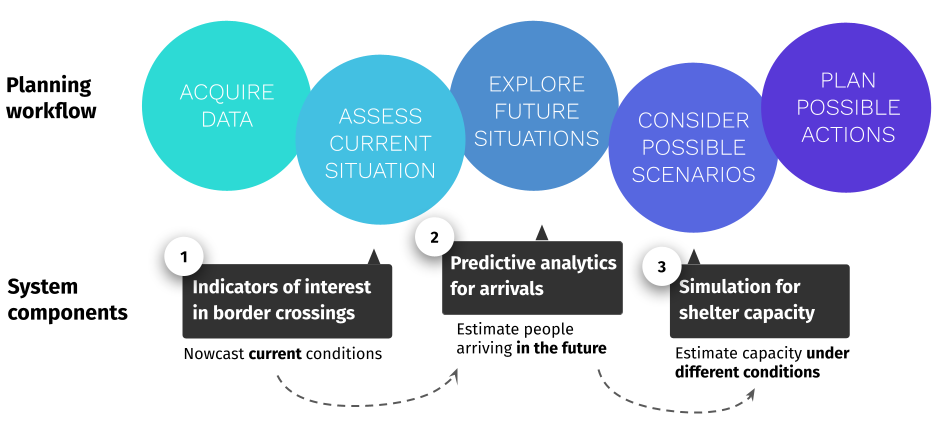}
    \caption{Illustration of where and how the three components of the {tool} aim to support the contingency planning process.}
    \label{fig:tool_overview}
\end{figure}

This {tool} aims to support humanitarian agencies by providing input into the contingency planning process for new migrant and refugee arrivals. {Contingency planning is a process by which UNHCR operations prepare for refugee emergencies. Specifically, a contingency plan is defined as ``a concise document that outlines the interagency response strategy with actions to be taken, by whom, where, and with what immediately available resources during the initial emergency response (first 3 months), should a specific risk scenario occur and once pre-identified activating triggers are met'' \cite{unhcr_ppre_2015}. Contingency planning is typically preceded by a risk analysis in order to identify these potential future scenarios and assess their likelihood. 

The situational analysis tool\footnote{Access to the {tool} is restricted due to the sensitivity of the data, but a public microsite is available at: \url{https://brazil-venezuela-flows.unglobalpulse.net/}.} described here is designed to support this process through three core components:} 
(1) \textbf{a collection of  “nowcast” data} which supplements operational data feeds with additional (near-)real-time potential indicators of interest in crossing the border; (2) \textbf{predictive models} to estimate future arrival levels according to different {socioeconomic} indicators and past border crossings; and 
(3) \textbf{an interactive simulation tool},  {which consists of a simple mathematical computation that models }how people move through the different stages of border crossing (e.g., arrival, registration, shelter, relocation) under various assumptions, with an emphasis on understanding how this will affect capacity in UNHCR's shelter system. The components of the {tool} are illustrated in Figure \ref{fig:tool_overview}. All three parts are designed to be used together in order to improve situational awareness, while avoiding over-reliance on any one specific (and potentially biased or unreliable) dataset or prediction.

\subsection{Part 1: Nowcasting current conditions} The first component {of the tool} is a {visualization of a set of} indicators that aim to track conditions related to border crossing activity, identify large numbers of people traveling to (or planning to travel to) the border, and potentially estimate the number of people that have already crossed the border. {Open-source} data is collected on a variety of factors expected to be correlated with border crossings, including: conflict/protest events and fatalities from the Armed Conflict Location and Event Dataset \cite{raleigh_introducing_2010}; community mobility patterns \cite{google_covid-19_2022};
 oil prices, consumer price indices, and exchange rates \cite{us_energy_information_administration_spot_2022, food_and_agriculture_organization_of_the_united_nations_consumer_2021,investingcom_usd_2022}; and COVID-19 case rates \cite{ritchie_coronavirus_2020}. {Border crossing intent is captured by} internet search activity for border regions and related topics \cite{google_google_2021}. {UNHCR's protection team regularly monitors community-based social media groups where refugees exchange information; therefore, we have also explored social media monitoring using an out-of-the-box analysis tool, as well as the use of social media audience estimates along the lines of \cite{palotti_monitoring_2020}}. 

\subsection{Part 2: Predictive analytics for future movements under uncertainty}
\begin{figure}
    \vspace{0.5cm}
    \centering
~\\~\\~\\~\\~\\
    \includegraphics[width=5in]{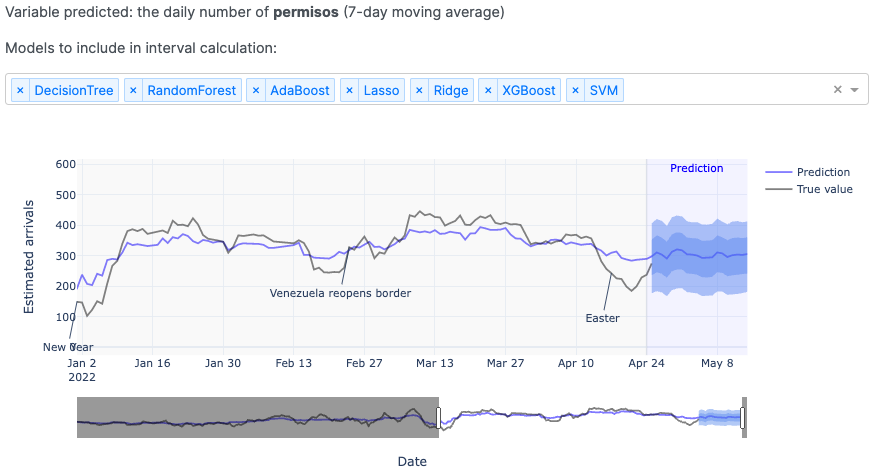}
    \caption{The {\textbf{predictive models} offer} a forecast of future trends in forced displacement from Venezuela to Brazil. The chart above shows consolidated output from all of the predictive models we trained. The ``true'' trendline shows the daily number of entry permits (7-day moving average) which has been fed into the models as the ``ground truth''. The ``prediction'' line represents the estimated number of arrivals according to an ensemble (weighted average) of all models. The weights for the average were determined by the models' performance on an unseen test dataset.}
    \label{fig:prediction}
\end{figure}

The second component of the {tool} is a set of machine learning models for predicting arrivals based on past border crossings. In our application setting, the target variable is the 7-day moving average of arrivals,\footnote{This average was taken to smooth over seasonality according to the day of the week.} 
predicted 30 days in advance. {Since there is no definitive measure of arrivals, we have explored the use of three different proxy variables: the number of official border crossings (the best indicator of crossings in the pre-COVID period); the number of new registrations in UNHCR's case management system (the best indicator of informal crossings when the border was closed); and the number of entry permits issued by the Brazilian government (the best indicator of crossings since Brazil has formally reopened the border).} The feature variables include indicators drawn from the different social, economic, and other non-traditional sources of data described above, as well as manually engineered features such as binary indicators for holidays 
as well as key events.
\footnote{For example, we added indicators for the Christmas holiday week; the Easter holiday week; school summer holidays; days on which buses were scheduled to travel to the border; dates on which the border was closed in one or both directions; and a set of notable dates identified during a manual review of news articles and related sources by the team.}

{Currently, the models are trained on data from July 26th, 2021 to January 31, 2022 and evaluated on data from February 1, 2022 to April 24, 2022.\footnote{The start date of the training period is approximately one month after Brazil reopened its borders to Venezuelans, which coincides with the start of our
data on entry permits issued. Since we predict arrivals 30 days in advance and use lagged arrivals in the prediction, this is the first date for which the full
feature set is available.}} {Generally speaking, we follow a modeling approach similar to that outlined in the case study of \cite{hoffmann_pham_predictive_2022}; we plan to expand this current case study into a full paper with more details on the predictive models and our methodology in the future.}
{Given the limited size of the training dataset, 
we opt for simplicity and fit a set of }standard regression models including lasso and ridge regressions \cite{tibshirani_regression_1996, hoerl_ridge_1970}, decision trees \cite{breiman_classification_2017}, random forests \cite{breiman_random_2001},  AdaBoost \cite{freund_short_1999}, XGBoost \cite{chen_xgboost_2016}, and support vector machines (SVM) \cite{cortes_support-vector_1995}. {Model parameters were selected via ten-fold cross validation using a blocking time-series split, in order to minimize mean squared error (MSE). These models were refit on the full dataset (July 26, 2021 - April 24, 2022) before making predictions.  While the predictions are noisy, we find that all but two models outperform a default strategy of predicting the historical mean number of arrivals in the training period when ranked by Root Mean Square Error (RMSE), and all models outperform the historical mean when ranked by Mean Absolute Error (MAE). }

{In an effort to produce more stable predictions and present a simple result to operational end-users,} forecasts are averaged into a single ensemble, where the weights in the average are determined by the RMSE of each individual component model. Prediction intervals for individual models are estimated using a bootstrapping approach \cite{efron_bootstrap_1979} and they are subsequently aggregated to produce prediction intervals for the ensemble as well. This approach communicates the fact that predictions are approximate, while also limiting the overall volume of conflicting information conveyed to the end user (relative to, for example, showing the predictions produced by each individual model). A snapshot of the predictions rendered by the tool is shown in Figure \ref{fig:prediction}.

Exploration of alternative predictive models to improve performance is an area of ongoing work, including gravity models \cite{anderson_gravity_2010}, long short-term memory (LSTM) models \cite{hochreiter_long_1997}, {and standard autoregressive time series (i.e., ARIMA) models. We are also investigating whether modeling arrivals at the subnational level (by origin state in Venezuela) might improve predictive performance and yield deeper insights into drivers of migration.}

\subsection{Part 3: Simulation for shelter capacity}

\begin{figure}
    \centering
    \includegraphics[width=6in]{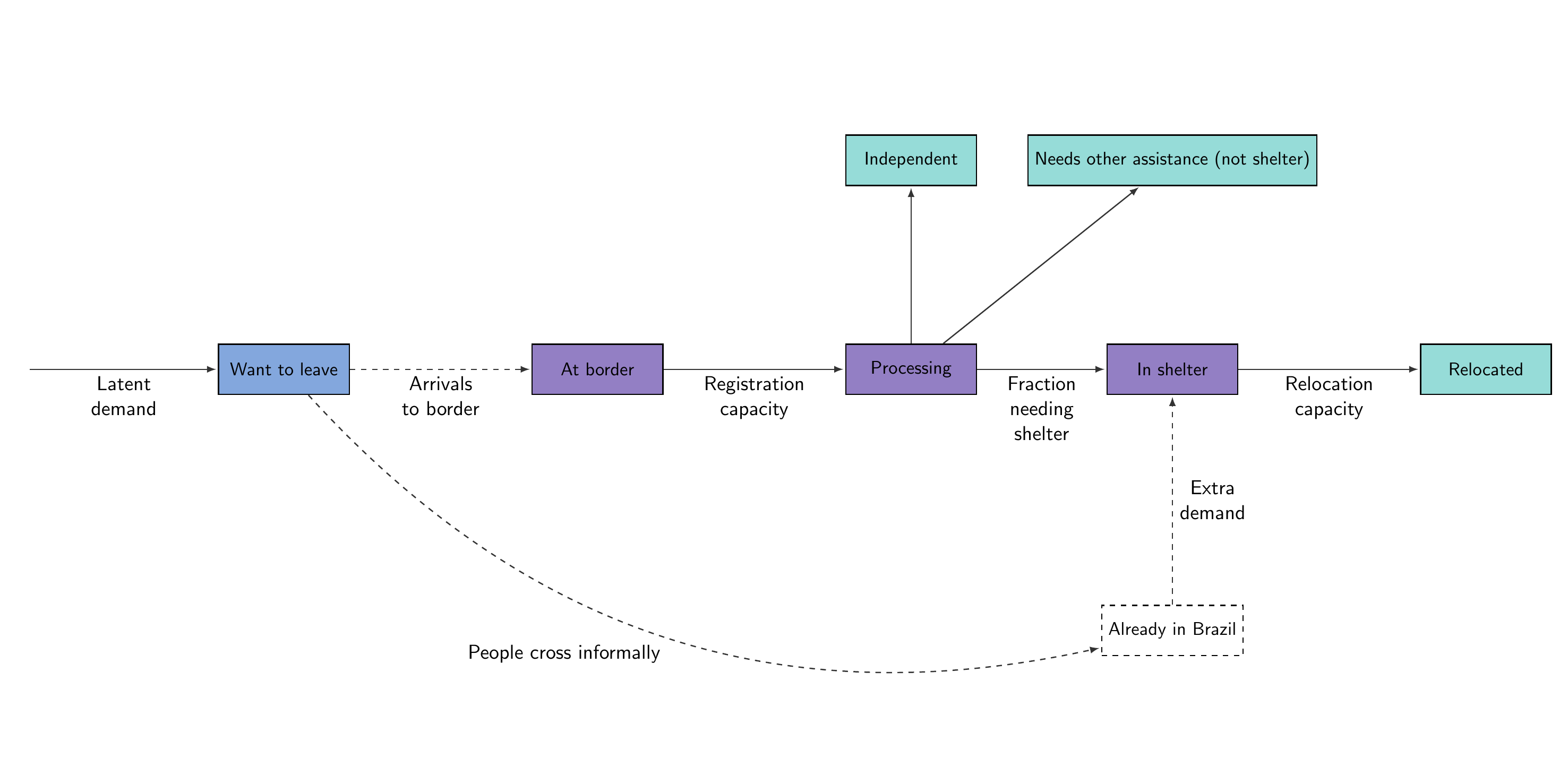}
    \caption{The \textbf{border crossing model} used in the simulation. The boxes represent the various stages of border crossing (i.e. wanting to leave, arriving at the border, undergoing processing, residing at a shelter, and being relocated to the interior of Brazil). The model flows from left to right. The text below each arrow indicates the parameter that determines how many people move between the two associated stages. For instance, ``registration capacity'' determines how many people are moved from the ``at border'' stage to the ``processing'' stage.}
    \label{fig:model}
\end{figure}

The final component of the {situational analysis tool is a simulation that} illustrates what might happen under different scenarios and assumptions.  The simulation is based on a {simple} compartmental model that captures the different stages of the border crossing (see Figure \ref{fig:model}), {which was designed through consultations with operational staff on the ground}. {At each time step, individuals are added to the ``want to leave'' pool on the left-hand side of the model; individuals can advance one stage forward in every subsequent time period, subject to the constraints imposed by the model assumptions. Where the number of users flowing into a given stage is greater than the number of users flowing out (for example, if there are 500 arrivals to the border per day but the registration capacity is only 300 people per day), bottlenecks will form.} 

\begin{figure}
    \vspace{0.5cm}
    \centering
    \includegraphics[width=6in]{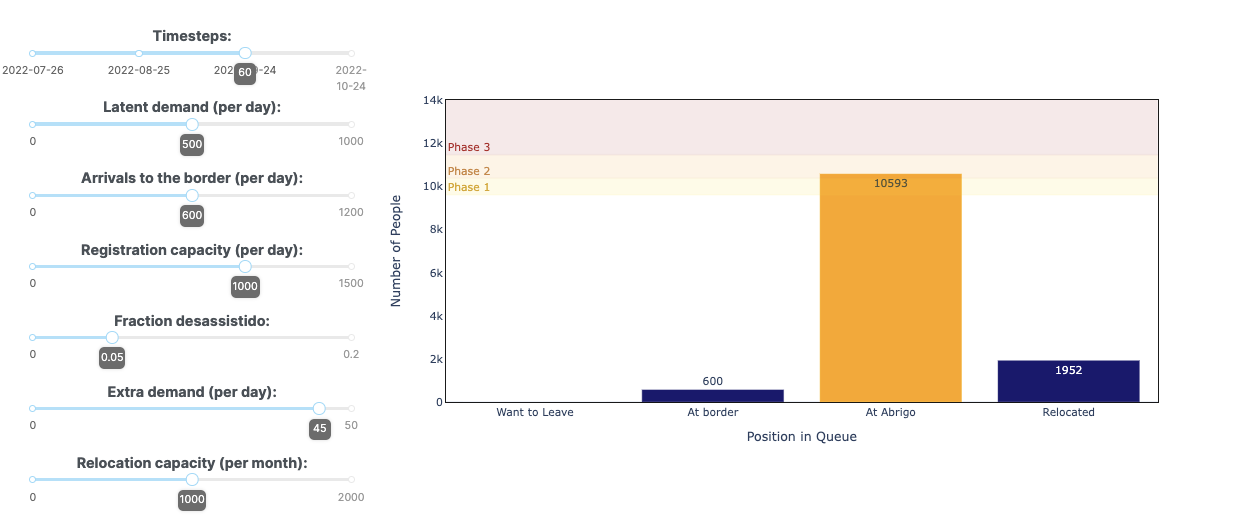}
   \caption{The \textbf{simulation} illustrates what might happen under different scenarios and certain assumptions. Users can move the sliders to vary several parameters related to rates of border crossing and relocation, in order to view the resulting number of people in each stage according to those parameters.}
   \label{fig:simulation}
\end{figure}

As shown in Figure 
\ref{fig:simulation}, the planner 
can {drag sliders to} explore varying assumptions about {latent demand for crossing the border, arrivals rates {at the border}, {the UNHCR operation's registration capacity}, the fraction of people with special needs (desassistidos), extra requests for shelter from people already in Brazil, and the operation's ability to relocate displaced people from temporary emergency shelters to residence in other parts of Brazil.} {The ranges of the sliders were selected based on past experience of the plausible values.} {When a slider is changed, the visualization displays} the estimated resulting number of people {at different stages of the border crossing at a given point in time. The design of the visualization emphasizes} the number of people in the shelter system, {which} allows the operational end-user to explore what conditions will lead to shelters reaching or exceeding capacity (and by how much). {This helps them to consider} potentially expanding capacity in advance of new arrivals, and to plan for additional needs such as COVID testing and vaccination. 

\begin{figure}
    \centering
    \includegraphics[width=6in]{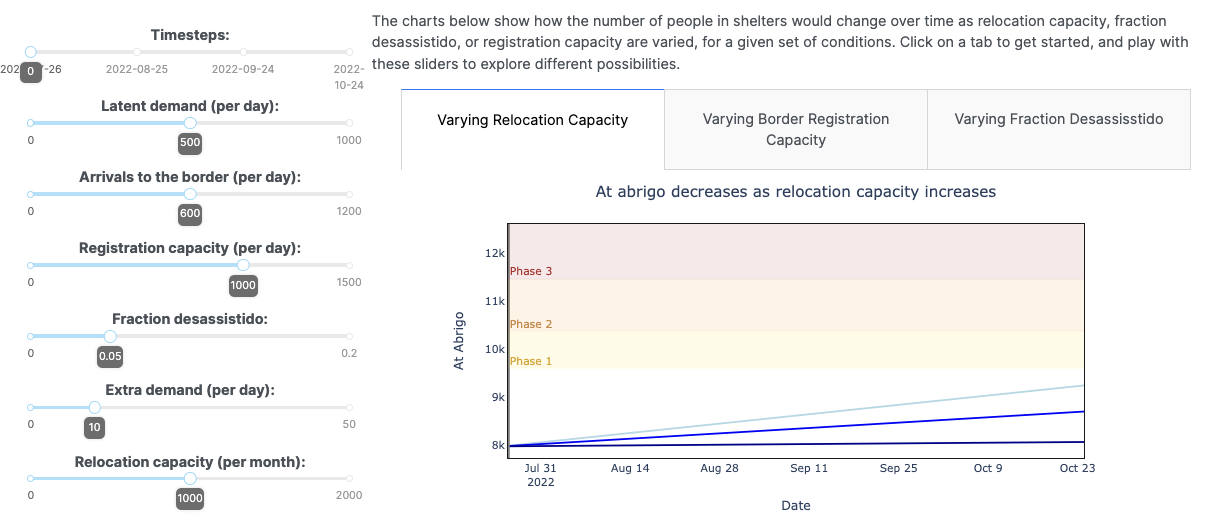}
   \caption{The \textbf{sensitivity analysis} illustrates how the number of people in shelters would change over time when only one of the parameters is varied (the other parameters remain constant and are set using the sliders). The graph here depicts the decrease in the number of people in the shelters (abrigos) when relocation capacity increases.}
   \label{fig:sensitivity}
\end{figure}

The simulation is supplemented by a sensitivity analysis that deliberately guides the decision maker through variations in the input parameters (see Figure \ref{fig:sensitivity}). The sensitivity analysis allows for two types of exploration: examination of how variations in a parameter affect the system (1) over time; and (2) at a given point in time. Ongoing work examines strategies for giving sophisticated users more control over the sensitivity analysis, while concurrently presenting a simpler ``narrative view'' for users who have limited time to explore.

\section{Co-design and operational deployment}
Over the course of this project, we have applied a human-centered approach to data science. A design process was conducted with prospective users at UNHCR Brazil and the modeling team, involving recurring meetings as well as dedicated workshops to understand operational needs and the specifics of the operational context{, including how decision makers expect to receive information and how simulations have been shared with external actors}. One of the early design activities involved mapping the stages of the border crossing, shown in Figure \ref{fig:model}. As a result, the model developed through this process closely reflects the actual operation. We incorporate logic from the contingency plan into the outputs of the simulation, indicating the conditions under which different parts of the plan may be triggered.

The initial goal of this project was to evaluate the likelihood of different scenarios under consideration by UNHCR, each of which is associated with a different preparatory operational action. {While there was a clear need for predictions at the outset of the pandemic, there were} several challenges to producing accurate predictive models in this context,
including: limited amounts of historical training data; anomalies in behavior due to COVID-19 and associated policy changes; and the difficulty in disentangling the effects of different health, socioeconomic, and social drivers of movement. 
The three-part approach described above was designed with these limitations in mind. In particular, nowcast indicators are collected to capture different signals of movement, which operational users can compare against their own datasets in order to identify discrepancies or trends of interest. 
On the other hand, the predictive models are used to produce a set of possible and plausible ranges for future arrivals, which can then be explored in more depth using the shelter capacity simulation.
While the {tool} itself is designed to highlight the complexity and ambiguity in the situation at the border, it also incorporates features that allow decision makers to explore particular aspects of this uncertainty.

{The primary applications of the tool are envisioned to be contingency planning, shelter allocation, and the prepositioning of humanitarian assistance. Thus far, UNHCR has used the tool for planning purposes and to support efforts to allocate resources in a more efficient and timely manner. During the border closure period, the simulation was the most actively used component of the tool, since it allows exploration without requiring training data from comparable periods, and it could be easily shared with external stakeholders since it does not use sensitive datasets. As additional historical training data becomes available from the period after the reopening of the border, we expect that the performance of the predictive models will improve and that these models will play a larger role. We also plan to begin exploring patterns of returns to Venezuela.}

To date, {the tool has had two key operational impacts}. 
First, in the particular operational context of Brazil, the opening of new emergency shelters is a political decision that is taken by Operação Acolhida (Operation Welcome), the federal government’s response to the influx of refugees and migrants from Venezuela. {The operation} is focused on reception and documentation, humanitarian assistance (including shelter), and voluntary internal relocation to other regions of Brazil \cite{government_of_brazil_sobre_2022}. By including additional evidence in the operation’s contingency plan for increased population flows, UNHCR and its partners were able to move to a technical negotiation. Second, over the course of this project, UNHCR learned to identify seasonal movement patterns and adjust accordingly. Thus, UNHCR provided public authorities with evidence to introduce adjustments in the shelter emergency response reflecting actual/evolving needs. {In the future, we plan to facilitate such applications by continuing to make the models and simulation more understandable and user-friendly, particularly when shared with audiences such as senior decision-makers, policymakers, and nontechnical staff.}
 
\section{Discussion and future work}
Now that COVID-related entry restrictions have been lifted and movement patterns have begun to establish a “new normal”, we plan to continue to improve the predictive models in order to support ongoing operations. While creating reliable forecasts in the face of emerging crises (such as the COVID-19 pandemic) is difficult, adopting a flexible approach that incorporates different models and perspectives may provide insights into the situation and possible outcomes, while highlighting ongoing uncertainty and avoiding overconfidence in one set of forecasts. 

In the future, we would like to extend the approach developed here to other forced displacement contexts. As noted above, previous work at UNHCR has explored the use of predictive analytics in Somalia \cite{unhcr_innovation_predictive_2019} and the Sahel \cite{adelphi_research_anticipating_2021}, and there is the potential for additional applications in emerging situations such as Ukraine. As the sophistication and accuracy of these models improves, they could be expanded to a broad range of humanitarian and/or disaster-related settings. For example, in addition to acting as planning and advocacy tools, displacement models can potentially help to shed light onto the drivers and/or early warning signs of displacement, with implications for conflict prevention, risk management, and building resilience. Through forecast-based financing \cite{ifrc_forecast-based_2022}, such models could even be used to proactively target assistance before the advent of a crisis (for example, in the context of climate-induced displacement). They could also be broadened to estimating needs beyond shelters or camps, with the goal of anticipating demand for job opportunities, food assistance, or housing in host communities.

\begin{acks}
United Nations Global Pulse is supported by the Governments of Sweden and Canada. {We thank Patricia Angkiriwang and Sherry Feng for their contributions to the design of the simulation tool.}
\end{acks}

\bibliographystyle{ACM-Reference-Format}
\bibliography{Venezuela}

\end{document}